\documentclass[aps,prl,reprint,superscriptaddress]{revtex4-1}

\usepackage[T1]{fontenc}
\usepackage{amsmath}
\usepackage{amssymb}
\usepackage{graphicx}
\usepackage{braket}
\usepackage{color}

\newcommand{\KTP}{\ensuremath{\mathrm{KTiOPO_4}} }
\newcommand{\op}[1]{\ensuremath{\mathrm{\hat{#1}}}}
\newcommand{\chitwo}{\ensuremath{\chi^{(2)}} }
\newcommand{\gtwo}{\ensuremath{ \text{g}^{(2)}} }

\newcommand{\hc}{\textrm{h.\,c.}}
\newcommand{\unit}[2]{\ensuremath{#1\,\textrm{#2}}}
\newcommand{\func}[2]{#1\left(#2\right)}

\newcommand{\vac}{\ket{0}}

\newcommand{\Hpdc}{\op{H}_\textrm{PDC}}
\newcommand{\A}{\op{A}}
\newcommand{\B}{\op{B}}
\newcommand{\imagewidth}{0.5\columnwidth}
\newcommand{\fint}[1]{\int\!\!\!\mathrm{d}#1}

\begin{document}

\title{A bright, pulsed two-mode squeezer}

\author{Andreas Eckstein}
\email[]{andreas.eckstein@mpl.mpg.de}
\author{Andreas Christ}
\affiliation{Max Planck Institute for the Science of Light, G\"unther-Scharowsky-Str. 1, 91054 Erlangen, Germany}


\author{Peter J. Mosley}
\affiliation{University of Bath, BA2 7AY, Bath UK}

\author{Christine Silberhorn}
\affiliation{University of Paderborn, Warburgerstr. 100, 33098 Paderborn, Germany\\Max Planck Institute for the Science of Light, G\"unther-Scharowsky-Str. 1, 91054 Erlangen, Germany}

\date{\today}

\begin{abstract}
We report the realization of a bright ultrafast two-mode squeezer based on type II parametric downconversion (PDC) in periodically poled \KTP (PP-KTP) waveguides. It produces a pulsed two-mode squeezed vacuum state: a photon-number entangled pair of truly single-mode pulses or, in terms of continuous variables quantum optics, a pulsed, single mode Einstein-Podolsky-Rosen (EPR) state in the telecom regime. We prove the single mode character of our source by measuring its $\gtwo$ correlation function and demonstrate a mean photon number of up to 2.5 per pulse, equivalent to 11dB of two-mode squeezing.
\end{abstract}

\maketitle

\noindent The main obstacle to the real-world deployment of wide area quantum communication networks is the limited distance of guaranteed security between communication partners. In order overcome it, quantum repeaters\cite{Briegel98} are needed to counter the security-degrading effects of transmission losses. For continuous variable (CV) quantum communication, these protocols heavily rely on the concatenation of non-Gaussian states and squeezed Gaussian states\cite{Eisert02,Giedke02}, namely EPR states produced by parametric downconversion (PDC) combined with photon counting\cite{Opatrny00}. In general though, PDC does not produce single mode but multimode EPR states, requiring additional post-processing for optimal fidelity. Their multimode structure is intrinsic to their generation process\cite{Martinelli03}, and only direct manipulation of that process allows for the production of single mode states.

For the generation of photon pairs, PDC sources have become an established standard: Inside a \chitwo-nonlinear medium, a pump photon decays into one signal and one idler photon. Recent works have shown that PDC source engineering\cite{Grice01,URen05} is capable of producing spectrally separable two-photon states $\ket{1}_s \otimes \ket{1}_i$, allowing for the preparation pure heralded single photons\cite{Mosley08}. Going beyond the single photon pair approximation, PDC in general can be understood as a source of squeezed states of light\cite{Wu86,Wasilewski06}. First observed by Slusher et al.\cite{Slusher85} in 1985, squeezed states originally garnered interest for the noise reduction in their quadrature observables $\op{X}, \op{Y}$ below the classical shot noise level, applicable in quantum-enhanced interferometry\cite{Caves81}.
With the availability of mode locked lasers, multimode\cite{Opatrny02} pulsed squeezed states\cite{Slusher87} became accessible. Measuring with detectors incapable of resolving this multimode structure, such as avalanche photo diodes (APD) implementing non-Gaussian operations\cite{Browne03}, introduces mixedness which degrades the quantum features of the state\cite{Rohde07}. Until now, PDC experiments rely on spatial\cite{Mosley08} or narrow spectral\cite{Tapster98} filtering of multimode\cite{Opatrny02,Wasilewski06} squeezers to approximate single mode bi-photonic states, with severe loss of source brightness.
In recent years, waveguide PDC sources\cite{Anderson95,Tanzilli02,Kanter02,Politi08} have become more and more popular as a means of achieving higher brightness in a single-pass configuration, as well as for their easy integrability into miniaturized quantum optical experiments.

In this Letter, we demonstrate a source of ultrafast single mode EPR states of unprecedenced brightness in the telecom wavelength regime. For low pump powers ($\braket{n}\ll 1$), it doubles as a source of pure heralded single photons\cite{Mosley08}. Utilizing a type II PDC process inside a single-mode PP-KTP waveguide as well as spectral engineering\cite{Grice01,URen05,Mosley08,Kuzucu08}, we are able to avoid narrow spectral or spatial filtering entirely, boosting source brightness considerably. The ultrafast, broadband nature of the pump beam prompts us to model the system in terms of broadband frequency modes\cite{Law00}. 
Our source emits pairs of spectrally broadband single mode pulses. This we corroborated by a measurement of spectral separability, a \gtwo measurement of one of the output arms to ensure the expected photon statistics, and a $\textrm{sinh}^2$ gain in mean photon number. The maxiumum mean photon number per pump pulse measured was 2.5, corresponding to 11dB of two-mode squeezing.

\begin{figure}[htb]
    \includegraphics[width=\linewidth]{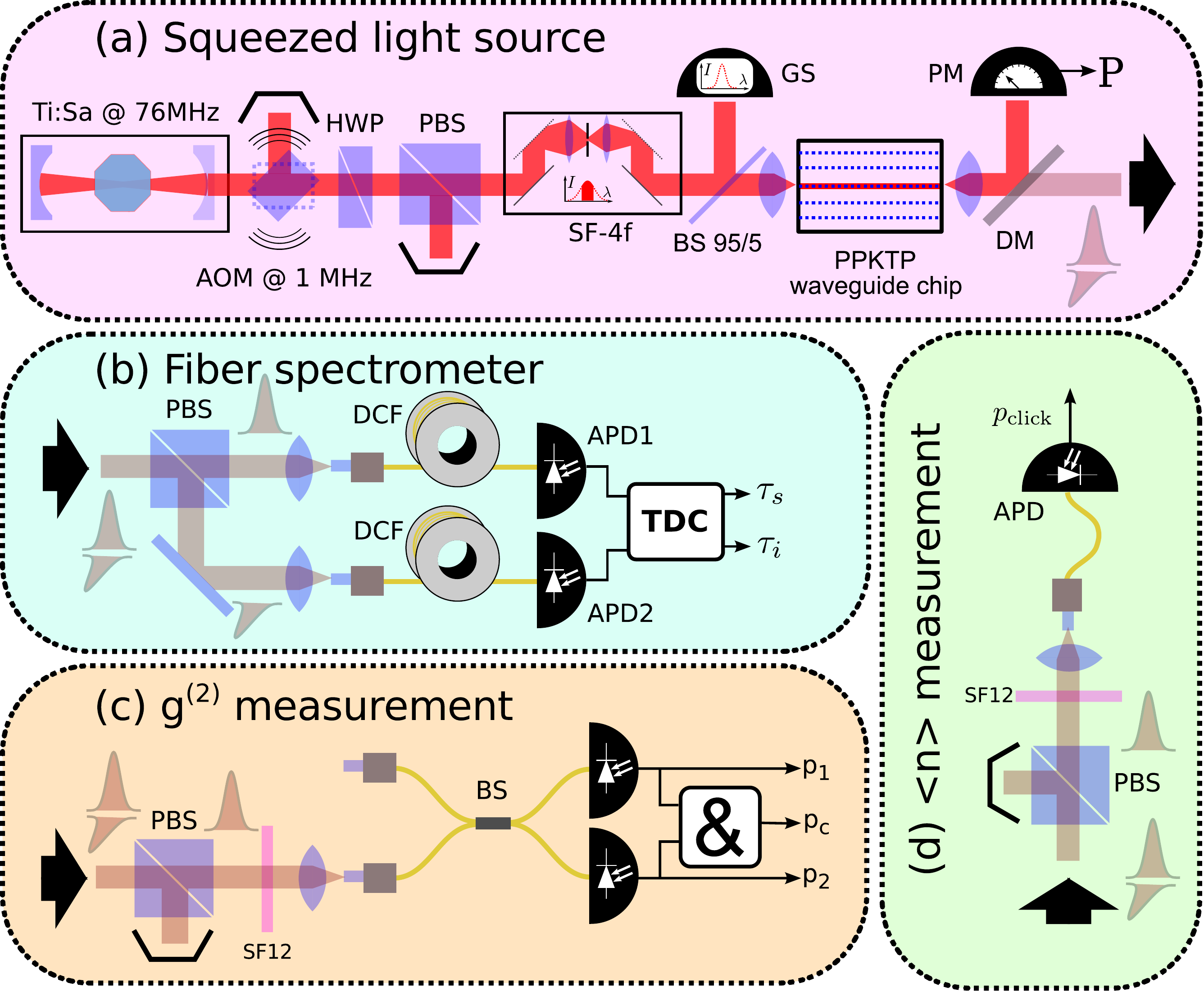}
    \caption{
    Experimental setup: (a) PP-KTP waveguide source of two-mode squeezed vacuum states. (b) Fiber spectrometer to measure the photon pair joint spectral intensity. (c) \gtwo measurement setup. (d) Setup to measure mean photon number $\left<n\right>$.
} 
    \label{fig:setup}
\end{figure}

It has been shown early on in the experimental exploration of squeezing that PDC produces squeezed vacuum states of light\cite{Wu86}. In photon number representation, a two-mode squeezed vacuum state or single mode EPR state has the form
\begin{align}
\ket{\psi}&=\op{S}_{a,b}\vac=e^{\imath \op{H}_{a,b}}\vac
=\sqrt{1-\left|\lambda\right|^2}\sum_n \lambda^n \ket{n,n}\label{eq:twomodesqueezedvac}
\end{align}

where $a$ and $b$ are two orthogonal modes, $\op{S}_{a,b}$ is the two-mode squeezing operator, and $\op{H}_{a,b}=\zeta \op{a}^\dagger \op{b}^\dagger + \hc$ is its effective Hamiltonian. It is a coherent superposition of strictly photon number correlated Fock states, and exhibits thermal photon statistics in both modes $a$ and $b$. The photon number correlation between both modes allows for heralding pure  single photons with binary detectors. However, the underlying bilinear effective Hamiltonian $\op{H}_{a,b}$ describes only a special case of PDC. 

In general, the effective PDC Hamiltonian has a richer spatio-spectral structure, but using a single mode waveguide allows us to restrict our analysis to one spatial mode, and we find
\begin{align}
\Hpdc=\zeta \fint \omega_1 \fint \omega_2 \func{f}{\omega_1, \omega_2} \op{a}^\dagger(\omega_1) \op{b}^\dagger(\omega_2) + \hc
\end{align}
which generates a generalized version of the two-mode squeezed vacuum in Eq. \ref{eq:twomodesqueezedvac} with spectrally correlated output beams. The coupling constant $\zeta$ determines the strength of this interaction, while spectral correlations between photons of the pairs produced are governed by the normalized joint spectral amplitude $\func{f}{\omega_1, \omega_2}$.

By applying a Schmidt decomposition to the joint amplitude\cite{Law00} $\func{f}{\omega_1, \omega_2}=\sum_k c_k\func{\varphi_k}{\omega_1}\func{\psi_k}{\omega_2}$, we obtain two orthonormal basis sets $\func{\varphi_k}{\omega_1}$ and $\func{\psi_k}{\omega_2}$ and a set of weighting coefficients $c_k$ with $\sum_k \left|c_k\right|^2=1$. Now the PDC Hamiltonian can be expressed in terms of broadband modes
\begin{align}
\Hpdc=\sum_k \op{H}_k = \zeta\sum_k c_k\left(\A_k^\dagger \B_k^\dagger + \A_k \B_k\right).
\end{align}
Each broadband mode operator $\A_k, \B_k$ describes a temporal pulse mode, or equivalently a ultrafast spectral mode. It is defined as superposition of monochromatic creation/annihilation operators $\hat{a}\left(\omega\right), \hat{b}\left(\omega\right)$ operators weighted with a function from the Schmidt basis: $\A_k^\dagger:=\int \mathrm d \omega \func{\varphi_k}{\omega} \op{a}^\dagger\left(\omega\right)$ and $\B_k^\dagger:=\int \mathrm d \omega \func{\psi_k}{\omega} \op{b}^\dagger\left(\omega\right)$. The effective Hamiltonians $\op{H}_k$ do not interact with each other (since $\left[\op{H}_k,\op{H}_l\right]=0$), and thus the PDC squeezing operator represents in fact an ensemble of independent two-mode squeezing operators $\op{S}_{a,b}=e^{\imath \Hpdc}=\op{S}_{A_0,B_0}\otimes \op{S}_{A_1,B_1} \otimes ...$ where the coefficients $c_k$ determine the relative strength of all squeezers as well as spectral correlation between signal and idler beams. This correlation is characterized by the source's effective mode number $K=\frac{1}{\sum_k \left|c_k\right|^4}$. For $c_0=1$ and all other $c_k=0$, $K$ assumes its minimum value of $1$, and the PDC process can be described as a two-mode squeezer according to Eq. \ref{eq:twomodesqueezedvac}.

In our waveguided source pumped by an ultrafast pulsed laser beam we can manipulate spectral correlations of the photon pair joint spectra, thus the coefficients $c_k$, and as a result minimize $K$ by simply adjusting the spectral width of the pump pulses \cite{Grice01,URen05}.

\begin{figure*}[htb]
    \includegraphics[width=\textwidth]{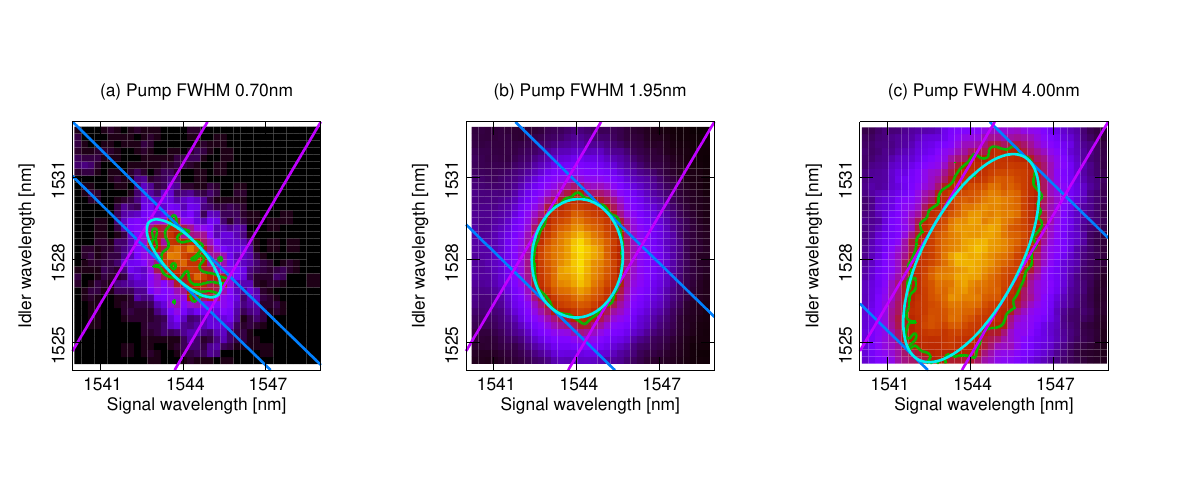}
    \vspace*{-1.75cm}
    \caption{Two-photon spectral intensities from setup \ref{fig:setup}(b) with pump width above, equal to and below photon pair separability width at \unit{1.95}{nm} FWHM. Green: 50\% intensity, violet: phasematching width, blue: pump width, bright blue: theo. 50\% intensity}\label{fig:jsi}
\end{figure*}

\begin{figure}[htb]
    \includegraphics[width=\imagewidth]{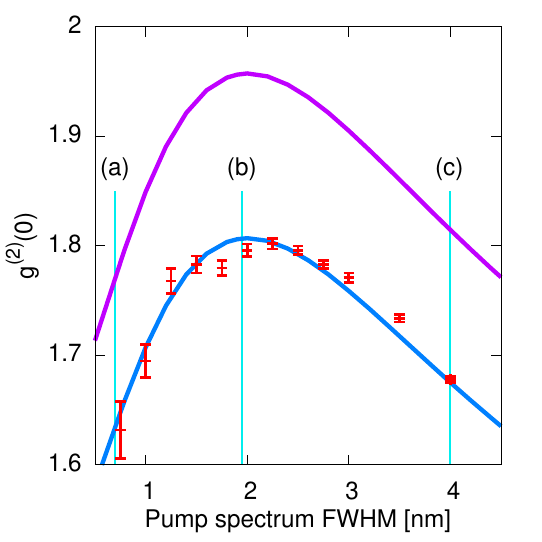}\includegraphics[width=\imagewidth]{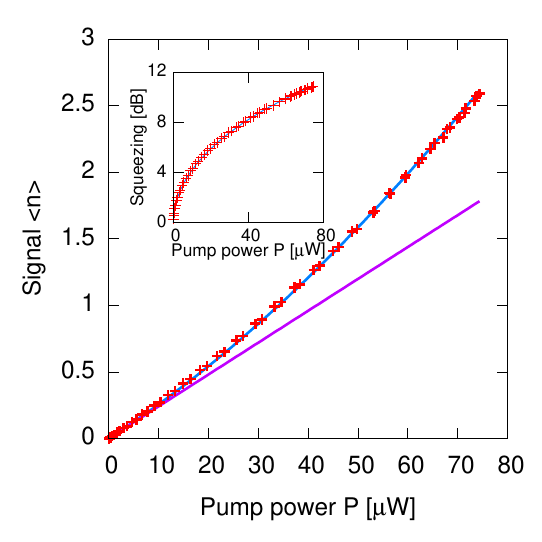}
    \caption{Left: \gtwo values from setup \ref{fig:setup}(c) (red) with theory curve (blue) and background corrected theory curve (vio    
    let); (a), (b) and (c) designate pump FWHM of the respective joint spectral intensities from Fig. \ref{fig:jsi}. Right: Mean photon number from setup \ref{fig:setup}(d) (red) with the theo. gain of a two-mode squeezer (blue) and the linear gain of a highly multimode squeezer (violet); inset: corresponding two-mode squeezing values.}\label{fig:g2_and_power}
\end{figure}

We verified this by measuring the joint spectral intensity of generated photon pairs at different spectral pump widths. The setup in Fig. \ref{fig:setup} (a) illustrates the PDC source: Ultrafast pump pulses at \unit{768}{nm} are prepared with a TiSa mode locked laser system, spectrally filtered with a variable bandpass filter 4f setup, and then used to pump a type II PDC process within the PP-KTP waveguide with a poling period of $104\mu m$. Its length is \unit{10}{mm} but an effective length of \unit{8}{mm} is used to correctly predict the measurement results in Figs. \ref{fig:jsi} and \ref{fig:g2_and_power}, since manufacturing imperfections in poling period and waveguide diameter lead to a widened phasematching distribution $\func{\Phi}{\omega_1,\omega_2}$ as if from a shorter waveguide. Central wavelengths of signal and idler beam were \unit{1544}{nm} and \unit{1528}{nm}, respectively. The generated photon pairs are analyzed in a fiber spectrometer\cite{Avenhaus09} (Fig. \ref{fig:setup}(b)): After separating signal and idler photons by polarization, they independently travel through long dispersive fibers, and are detected by a pair of idQuantique id201 avalanche photo diodes (APDs). Due to the chromatic dispersion of the fibers, the photons' group velocity and arrival time at the APDs depend on their wavelength. Thus we are able to determine the spectral intensity distribution of a stream of single photons from its arrival time spread. For a spectral pump FWHM of \unit{0.70}{nm}, \unit{1.95}{nm} and \unit{4.0}{nm}, we observe in Fig \ref{fig:jsi} negative spectral correlations, an uncorrelated spectrum, and positive spectral correlations between signal and idler photons, respectively. We have demonstrated control over spectral entanglement between signal and idler by filtering the pump spectrum, and found minimal spectral correlations of photon pairs around \unit{1.95}{nm} pump FWHM.


To prove the genuine two-mode squeezer character of our source, an uncorrelated joint spectral intensity is necessary but not sufficient. It is proportional to the modulus square of the complex joint amplitude $\left|\func{f}{\omega_1, \omega_2}\right|^2$ of the photon pair, so all phase information is lost in an intensity measurement. In order to detect phase entanglement between signal and idler, we need to measure an additional quantity sensitive to the source's mode number $K$, which is unity only in the absence of entanglement on the photon pair level, and larger otherwise.

The second order correlation function \gtwo can be used to discriminate between beams with thermal ($\gtwo=2$) and Poissonian photon statistics ($\gtwo=1$) from a PDC source\cite{Tapster98}. As has been noted above, type II PDC can in general be understood as an ensemble of two-mode squeezers, each of them emitting two beams with thermal photon statistics. In our waveguided type II setup, all broadband modes, $A_k$ or $B_k$, share one polarization mode, $a$ and $b$ or respectively. A detector with a spectral response function much wider than the characteristic width of the broadband modes cannot resolve them. It ``sees'' a convolution of the thermal photon statistics of all broadband modes, and in the limit of a large number of modes, this is a Poissonian distribution\cite{Avenhaus08}. But if there is only one mode per polarization to begin with (which is only true for a two-mode squeezer), the detector receives a thermal distribution of photon numbers. Therefore, with the assumption that PDC emits a pure state, we can infer from  $\gtwo=2$ measured in either output beam a two-mode squeezer source. Indeed, for low pump power and thus low coupling strength $\zeta$, we can find a simple connection between the \gtwo correlation function on the one hand, and the broadband mode structure of our source and the effecive mode number $K$ on the other: $\gtwo=1+\sum\left|c_k\right|^4=1+\frac{1}{K}$.

Fig. \ref{fig:setup} (c) illustrates the \gtwo measurement: Idler is discarded, and the signal beam is split by a 50/50 beamsplitter. Its output modes are fed into APDs, single (\(p_1, p_2\)) and coincidence (\(p_c\)) click probabilities for different spectral pump widths are recorded. When using binary detectors far from saturation, rather than intensity measurements, one finds 
$\gtwo=\frac{\left<\op{a}^\dagger \op{a}^\dagger \op{a} \op{a} \right>}{\left< \op{a}^\dagger \op{a} \right>^2}\approx\frac{p_c}{p_1 p_2}$
As has been demonstrated, frequency correlations between signal and idler beam and thus squeezer mode number can be controlled by manipulation of the spectral width of the PDC pump beam. In Fig. \ref{fig:g2_and_power} (left) measurement results show a maximum \gtwo value at \unit{1.95}{nm} pump FWHM, in accordance with Fig. \ref{fig:jsi}. When departing from the optimum pump width, \gtwo decreases as predicted. Due to residual background events from waveguide material fluorescence and detector dark counts, we obtain a maximum of $\gtwo=1.8$, and $\gtwo=1.95$ after background correction.
This highlights the next-to-perfect single mode EPR states our source emits, and the degree of control we exact over the mode number and photon statistics of the system. Note that the two-mode character is shown with respect to frequency as well as spatial degrees of freedom. Owing to the waveguide nature of our source, signal and idler beam occupy a single waveguide mode.

Nonlinear waveguides allow for dramatically higher source brightness when compared to bulk sources\cite{Fiorentino07}: Instead of coupling to a continuum of spatial modes, inside a waveguide structure the generated waves couple to a discrete spectrum, and ideally to just one mode, boosting self-seeding of the PDC process and greatly simplifying collection of the output light. At mean photon numbers of $\braket{n}\approx 1$ per mode we will be able to observe the superlinear gain of a two-mode squeezer $\textrm{sinh}^2\left(r\right)$ caused by self-seeding of signal and idler along the waveguide length, further corroborating our source's single mode character.
With a pump FWHM of \unit{1.95}{nm} producing separable photon pairs, we measured the mean photon number $\braket{n}\approx\frac{p_\textrm{click}}{\eta}$ of the signal beam (Fig. \ref{fig:setup} (d)) by recording the power dependent APD click probability $p_\textrm{click}$. For binary detectors far from saturation, this is proportional to $\braket n$, with an overall quantum efficiency $\eta$ of the setup. The source gain in Fig. \ref{fig:g2_and_power} (right) exhibits with increasing pump power the departure from the linear gain profile that would be expected for a highly multimode squeezer, while it is in very good agreement with the theoretical prediction for a two-mode squeezer gain. Mean photon numbers of up to 2.5 or equivalently \unit{11}{dB} of two-mode squeezing were achieved. For an optimized setup we observed an overall detection efficiency of 15\%, and for a specified APD quantum efficiency of 25\% at \unit{1550}{nm}, this makes a photon collection efficiency into single mode fiber of 60\%. Note that our waveguide output facet was not anti-reflection coated.


In conclusion, we have applied spectral engineering to a waveguided PDC source to create a bright, genuinely ultrafast pulsed two-mode squeezer in the telecom wavelength regime with mean photon number per pulse as high as 2.5, or \unit{11}{dB} of two-mode squeezing. It features near thermal photon statistics boasting a \gtwo value of 1.95 after background correction, or an effective mode number of $K=1.05$. A collection efficiency of 60\% into single mode fibers demonstrates the high spatial mode quality of our waveguide device and shows its potential for inclusion into integrated optical networks. Due to its true two-mode character and brightness,
we expect widespread adoption of our source in continuous variable quantum communication, where high squeezing values, purity and low-loss fiber transmission are prerequisite for efficient quantum cryptography\cite{Gottesmann01}, teleportation\cite{Furusawa98,Bowen04}, and ultimately entanglement distillation\cite{Opatrny00,Browne03} to overcome transmission losses in wide area quantum communication networks, a vital building block of quantum repeaters\cite{Briegel98}.

This work was supported by the EC under the grant agreement CORNER (FP7-ICT-213681).


\begin{thebibliography}{10}%
\makeatletter
\providecommand \@ifxundefined [1]{%
 \ifx #1\undefined \expandafter \@firstoftwo
 \else \expandafter \@secondoftwo
\fi
}%
\providecommand \@ifnum [1]{%
 \ifnum #1\expandafter \@firstoftwo
 \else \expandafter \@secondoftwo
\fi
}%
\providecommand \enquote [1]{``#1''}%
\providecommand \bibnamefont  [1]{#1}%
\providecommand \bibfnamefont [1]{#1}%
\providecommand \citenamefont [1]{#1}%
\providecommand\href[0]{\@sanitize\@href}%
\providecommand\@href[1]{\endgroup\@@startlink{#1}\endgroup\@@href}%
\providecommand\@@href[1]{#1\@@endlink}%
\providecommand \@sanitize [0]{\begingroup\catcode`\&12\catcode`\#12\relax}%
\@ifxundefined \pdfoutput {\@firstoftwo}{%
 \@ifnum{\z@=\pdfoutput}{\@firstoftwo}{\@secondoftwo}%
}{%
 \providecommand\@@startlink[1]{\leavevmode}%
 \providecommand\@@endlink[0]{}%
}{%
 \providecommand\@@startlink[1]{%
  \leavevmode
  \pdfstartlink
   attr{/Border[0 0 1 ]/H/I/C[0 1 1]}%
   user{/Subtype/Link/A<</Type/Action/S/URI/URI(#1)>>}%
  \relax
 }%
 \providecommand\@@endlink[0]{\pdfendlink}%
}%
\providecommand \url  [0]{\begingroup\@sanitize \@url }%
\providecommand \@url [1]{\endgroup\@href {#1}{\urlprefix}}%
\providecommand \urlprefix [0]{URL }%
\providecommand \Eprint[0]{\href }%
\@ifxundefined \urlstyle {%
  \providecommand \doi [1]{doi:\discretionary{}{}{}#1}%
}{%
  \providecommand \doi [0]{doi:\discretionary{}{}{}\begingroup
  \urlstyle{rm}\Url }%
}%
\providecommand \doibase [0]{http://dx.doi.org/}%
\providecommand \Doi[1]{\href{\doibase#1}}%
\providecommand \bibAnnote [3]{%
  \BibitemShut{#1}%
  \begin{quotation}\noindent
    \textsc{Key:}\ #2\\\textsc{Annotation:}\ #3%
  \end{quotation}%
}%
\providecommand \bibAnnoteFile [2]{%
  \IfFileExists{#2}{\bibAnnote {#1} {#2} {\input{#2}}}{}%
}%
\providecommand \typeout [0]{\immediate \write \m@ne }%
\providecommand \selectlanguage [0]{\@gobble}%
\providecommand \bibinfo [0]{\@secondoftwo}%
\providecommand \bibfield [0]{\@secondoftwo}%
\providecommand \translation [1]{[#1]}%
\providecommand \BibitemOpen[0]{}%
\providecommand \bibitemStop [0]{}%
\providecommand \bibitemNoStop [0]{.\EOS\space}%
\providecommand \EOS [0]{\spacefactor3000\relax}%
\providecommand \BibitemShut [1]{\csname bibitem#1\endcsname}%
\bibitem{Briegel98}%
  \BibitemOpen
  \bibfield{author}{%
  \bibinfo {author} {\bibfnamefont{H.-J.}\ \bibnamefont{Briegel}}, \bibinfo
  {author} {\bibfnamefont{W.}~\bibnamefont{D\"ur}}, \bibinfo {author}
  {\bibfnamefont{J.~I.}\ \bibnamefont{Cirac}},\ and\ \bibinfo {author}
  {\bibfnamefont{P.}~\bibnamefont{Zoller}},\ }%
  \bibfield{journal}{%
  \Doi{10.1103/PhysRevLett.81.5932}{\bibinfo {journal} {Phys. Rev. Lett.}}\ }%
  \textbf{\bibinfo {volume} {81}},\ \bibinfo {pages} {5932} (\bibinfo {month}
  {Dec}\ \bibinfo {year} {1998})%
  \bibAnnoteFile{NoStop}{Briegel98}%
\bibitem{Eisert02}%
  \BibitemOpen
  \bibfield{author}{%
  \bibinfo {author} {\bibfnamefont{J.}~\bibnamefont{Eisert}}, \bibinfo {author}
  {\bibfnamefont{S.}~\bibnamefont{Scheel}},\ and\ \bibinfo {author}
  {\bibfnamefont{M.~B.}\ \bibnamefont{Plenio}},\ }%
  \bibfield{journal}{%
  \Doi{10.1103/PhysRevLett.89.137903}{\bibinfo {journal} {Phys. Rev. Lett.}}\
  }%
  \textbf{\bibinfo {volume} {89}},\ \bibinfo {pages} {137903} (\bibinfo {month}
  {Sep}\ \bibinfo {year} {2002})%
  \bibAnnoteFile{NoStop}{Eisert02}%
\bibitem{Giedke02}%
  \BibitemOpen
  \bibfield{author}{%
  \bibinfo {author} {\bibfnamefont{G.}~\bibnamefont{Giedke}}\ and\ \bibinfo
  {author} {\bibfnamefont{J.}~\bibnamefont{Ignacio~Cirac}},\ }%
  \bibfield{journal}{%
  \Doi{10.1103/PhysRevA.66.032316}{\bibinfo {journal} {Phys. Rev. A}}\ }%
  \textbf{\bibinfo {volume} {66}},\ \bibinfo {pages} {032316} (\bibinfo {month}
  {Sep}\ \bibinfo {year} {2002})%
  \bibAnnoteFile{NoStop}{Giedke02}%
\bibitem{Opatrny00}%
  \BibitemOpen
  \bibfield{author}{%
  \bibinfo {author} {\bibfnamefont{T.}~\bibnamefont{Opatrn\'y}}, \bibinfo
  {author} {\bibfnamefont{G.}~\bibnamefont{Kurizki}},\ and\ \bibinfo {author}
  {\bibfnamefont{D.-G.}\ \bibnamefont{Welsch}},\ }%
  \bibfield{journal}{%
  \Doi{10.1103/PhysRevA.61.032302}{\bibinfo {journal} {Phys. Rev. A}}\ }%
  \textbf{\bibinfo {volume} {61}},\ \bibinfo {pages} {032302} (\bibinfo {month}
  {Feb}\ \bibinfo {year} {2000})%
  \bibAnnoteFile{NoStop}{Opatrny00}%
\bibitem{Martinelli03}%
  \BibitemOpen
  \bibfield{author}{%
  \bibinfo {author} {\bibfnamefont{M.}~\bibnamefont{Martinelli}}, \bibinfo
  {author} {\bibfnamefont{N.}~\bibnamefont{Treps}}, \bibinfo {author}
  {\bibfnamefont{S.}~\bibnamefont{Ducci}}, \bibinfo {author}
  {\bibfnamefont{S.}~\bibnamefont{Gigan}}, \bibinfo {author}
  {\bibfnamefont{A.}~\bibnamefont{Ma\^\i{}tre}},\ and\ \bibinfo {author}
  {\bibfnamefont{C.}~\bibnamefont{Fabre}},\ }%
  \bibfield{journal}{%
  \Doi{10.1103/PhysRevA.67.023808}{\bibinfo {journal} {Phys. Rev. A}}\ }%
  \textbf{\bibinfo {volume} {67}},\ \bibinfo {pages} {023808} (\bibinfo {month}
  {Feb}\ \bibinfo {year} {2003})%
  \bibAnnoteFile{NoStop}{Martinelli03}%
\bibitem{Grice01}%
  \BibitemOpen
  \bibfield{author}{%
  \bibinfo {author} {\bibfnamefont{W.~P.}\ \bibnamefont{Grice}}, \bibinfo
  {author} {\bibfnamefont{A.~B.}\ \bibnamefont{U'Ren}},\ and\ \bibinfo {author}
  {\bibfnamefont{I.~A.}\ \bibnamefont{Walmsley}},\ }%
  \bibfield{journal}{%
  \Doi{10.1103/PhysRevA.64.063815}{\bibinfo {journal} {Phys. Rev. A}}\ }%
  \textbf{\bibinfo {volume} {64}},\ \bibinfo {pages} {063815} (\bibinfo {month}
  {Nov}\ \bibinfo {year} {2001})%
  \bibAnnoteFile{NoStop}{Grice01}%
\bibitem{URen05}%
  \BibitemOpen
  \bibfield{author}{%
  \bibinfo {author} {\bibfnamefont{A.~B.}\ \bibnamefont{{U'Ren}}}, \bibinfo
  {author} {\bibfnamefont{C.}~\bibnamefont{Silberhorn}}, \bibinfo {author}
  {\bibfnamefont{K.}~\bibnamefont{Banaszek}}, \bibinfo {author}
  {\bibfnamefont{I.~A.}\ \bibnamefont{Walmsley}}, \bibinfo {author}
  {\bibfnamefont{R.}~\bibnamefont{Erdmann}}, \bibinfo {author}
  {\bibfnamefont{W.~P.}\ \bibnamefont{Grice}},\ and\ \bibinfo {author}
  {\bibfnamefont{M.~G.}\ \bibnamefont{Raymer}},\ }%
  \bibfield{journal}{%
  \bibinfo {journal} {Laser Physics}\ }%
  \textbf{\bibinfo {volume} {15}} (\bibinfo {year} {2005})%
  \bibAnnoteFile{NoStop}{URen05}%
\bibitem{Mosley08}%
  \BibitemOpen
  \bibfield{author}{%
  \bibinfo {author} {\bibfnamefont{P.~J.}\ \bibnamefont{Mosley}}, \bibinfo
  {author} {\bibfnamefont{J.~S.}\ \bibnamefont{Lundeen}}, \bibinfo {author}
  {\bibfnamefont{B.~J.}\ \bibnamefont{Smith}}, \bibinfo {author}
  {\bibfnamefont{P.}~\bibnamefont{Wasylczyk}}, \bibinfo {author}
  {\bibfnamefont{A.~B.}\ \bibnamefont{U'Ren}}, \bibinfo {author}
  {\bibfnamefont{C.}~\bibnamefont{Silberhorn}},\ and\ \bibinfo {author}
  {\bibfnamefont{I.~A.}\ \bibnamefont{Walmsley}},\ }%
  \bibfield{journal}{%
  \Doi{10.1103/PhysRevLett.100.133601}{\bibinfo {journal} {Phys. Rev. Lett.}}\
  }%
  \textbf{\bibinfo {volume} {100}},\ \bibinfo {eid} {133601} (\bibinfo {year}
  {2008})%
  \bibAnnoteFile{NoStop}{Mosley08}%
\bibitem{Wu86}%
  \BibitemOpen
  \bibfield{author}{%
  \bibinfo {author} {\bibfnamefont{L.}~\bibnamefont{Wu}}, \bibinfo {author}
  {\bibfnamefont{H.~J.}\ \bibnamefont{Kimble}}, \bibinfo {author}
  {\bibfnamefont{J.~L.}\ \bibnamefont{Hall}},\ and\ \bibinfo {author}
  {\bibfnamefont{H.}~\bibnamefont{Wu}},\ }%
  \bibfield{journal}{%
  \Doi{10.1103/PhysRevLett.57.2520}{\bibinfo {journal} {Phys. Rev. Lett.}}\ }%
  \textbf{\bibinfo {volume} {57}},\ \bibinfo {pages} {2520} (\bibinfo {month}
  {Nov.}\ \bibinfo {year} {1986})%
  \bibAnnoteFile{NoStop}{Wu86}%
\bibitem{Wasilewski06}%
  \BibitemOpen
  \bibfield{author}{%
  \bibinfo {author} {\bibfnamefont{W.}~\bibnamefont{Wasilewski}}, \bibinfo
  {author} {\bibfnamefont{A.~I.}\ \bibnamefont{Lvovsky}}, \bibinfo {author}
  {\bibfnamefont{K.}~\bibnamefont{Banaszek}},\ and\ \bibinfo {author}
  {\bibfnamefont{C.}~\bibnamefont{Radzewicz}},\ }%
  \bibfield{journal}{%
  \Doi{10.1103/PhysRevA.73.063819}{\bibinfo {journal} {Phys. Rev. A}}\ }%
  \textbf{\bibinfo {volume} {73}},\ \bibinfo {pages} {063819} (\bibinfo {month}
  {Jun}\ \bibinfo {year} {2006})%
  \bibAnnoteFile{NoStop}{Wasilewski06}%
\bibitem{Slusher85}%
  \BibitemOpen
  \bibfield{author}{%
  \bibinfo {author} {\bibfnamefont{R.~E.}\ \bibnamefont{Slusher}}, \bibinfo
  {author} {\bibfnamefont{L.~W.}\ \bibnamefont{Hollberg}}, \bibinfo {author}
  {\bibfnamefont{B.}~\bibnamefont{Yurke}}, \bibinfo {author}
  {\bibfnamefont{J.~C.}\ \bibnamefont{Mertz}},\ and\ \bibinfo {author}
  {\bibfnamefont{J.~F.}\ \bibnamefont{Valley}},\ }%
  \bibfield{journal}{%
  \Doi{10.1103/PhysRevLett.55.2409}{\bibinfo {journal} {Phys. Rev. Lett.}}\ }%
  \textbf{\bibinfo {volume} {55}},\ \bibinfo {pages} {2409} (\bibinfo {month}
  {Nov.}\ \bibinfo {year} {1985})%
  \bibAnnoteFile{NoStop}{Slusher85}%
\bibitem{Caves81}%
  \BibitemOpen
  \bibfield{author}{%
  \bibinfo {author} {\bibfnamefont{C.~M.}\ \bibnamefont{Caves}},\ }%
  \bibfield{journal}{%
  \Doi{10.1103/PhysRevD.23.1693}{\bibinfo {journal} {Phys. Rev. D}}\ }%
  \textbf{\bibinfo {volume} {23}},\ \bibinfo {pages} {1693} (\bibinfo {month}
  {Apr}\ \bibinfo {year} {1981})%
  \bibAnnoteFile{NoStop}{Caves81}%
\bibitem{Opatrny02}%
  \BibitemOpen
  \bibfield{author}{%
  \bibinfo {author} {\bibfnamefont{T.}~\bibnamefont{Opatrn\'y}}, \bibinfo
  {author} {\bibfnamefont{N.}~\bibnamefont{Korolkova}},\ and\ \bibinfo {author}
  {\bibfnamefont{G.}~\bibnamefont{Leuchs}},\ }%
  \bibfield{journal}{%
  \Doi{10.1103/PhysRevA.66.053813}{\bibinfo {journal} {Phys. Rev. A}}\ }%
  \textbf{\bibinfo {volume} {66}},\ \bibinfo {pages} {053813} (\bibinfo {month}
  {Nov}\ \bibinfo {year} {2002})%
  \bibAnnoteFile{NoStop}{Opatrny02}%
\bibitem{Slusher87}%
  \BibitemOpen
  \bibfield{author}{%
  \bibinfo {author} {\bibfnamefont{R.~E.}\ \bibnamefont{Slusher}}, \bibinfo
  {author} {\bibfnamefont{P.}~\bibnamefont{Grangier}}, \bibinfo {author}
  {\bibfnamefont{A.}~\bibnamefont{LaPorta}}, \bibinfo {author}
  {\bibfnamefont{B.}~\bibnamefont{Yurke}},\ and\ \bibinfo {author}
  {\bibfnamefont{M.~J.}\ \bibnamefont{Potasek}},\ }%
  \bibfield{journal}{%
  \Doi{10.1103/PhysRevLett.59.2566}{\bibinfo {journal} {Phys. Rev. Lett.}}\ }%
  \textbf{\bibinfo {volume} {59}},\ \bibinfo {pages} {2566} (\bibinfo {month}
  {Nov}\ \bibinfo {year} {1987})%
  \bibAnnoteFile{NoStop}{Slusher87}%
\bibitem{Browne03}%
  \BibitemOpen
  \bibfield{author}{%
  \bibinfo {author} {\bibfnamefont{D.~E.}\ \bibnamefont{Browne}}, \bibinfo
  {author} {\bibfnamefont{J.}~\bibnamefont{Eisert}}, \bibinfo {author}
  {\bibfnamefont{S.}~\bibnamefont{Scheel}},\ and\ \bibinfo {author}
  {\bibfnamefont{M.~B.}\ \bibnamefont{Plenio}},\ }%
  \bibfield{journal}{%
  \Doi{10.1103/PhysRevA.67.062320}{\bibinfo {journal} {Phys. Rev. A}}\ }%
  \textbf{\bibinfo {volume} {67}},\ \bibinfo {pages} {062320} (\bibinfo {month}
  {Jun}\ \bibinfo {year} {2003})%
  \bibAnnoteFile{NoStop}{Browne03}%
\bibitem{Rohde07}%
  \BibitemOpen
  \bibfield{author}{%
  \bibinfo {author} {\bibfnamefont{P.~P.}\ \bibnamefont{Rohde}}, \bibinfo
  {author} {\bibfnamefont{W.}~\bibnamefont{Mauerer}},\ and\ \bibinfo {author}
  {\bibfnamefont{C.}~\bibnamefont{Silberhorn}},\ }%
  \bibfield{journal}{%
  \bibinfo {journal} {New J. Phys.}\ }%
  \textbf{\bibinfo {volume} {9}},\ \bibinfo {pages} {91} (\bibinfo {year}
  {2007})%
  \bibAnnoteFile{NoStop}{Rohde07}%
\bibitem{Tapster98}%
  \BibitemOpen
  \bibfield{author}{%
  \bibinfo {author} {\bibfnamefont{P.~R.}\ \bibnamefont{Tapster}}\ and\
  \bibinfo {author} {\bibfnamefont{J.~G.}\ \bibnamefont{Rarity}},\ }%
  \bibfield{journal}{%
  \Doi{10.1080/09500349808231917}{\bibinfo {journal} {J. Mod. Opt}}\ }%
  \textbf{\bibinfo {volume} {45}},\ \bibinfo {pages} {595} (\bibinfo {year}
  {1998}),\ ISSN \bibinfo {issn} {0950-0340}%
  \bibAnnoteFile{NoStop}{Tapster98}%
\bibitem{Anderson95}%
  \BibitemOpen
  \bibfield{author}{%
  \bibinfo {author} {\bibfnamefont{M.~E.}\ \bibnamefont{Anderson}}, \bibinfo
  {author} {\bibfnamefont{M.}~\bibnamefont{Beck}}, \bibinfo {author}
  {\bibfnamefont{M.~G.}\ \bibnamefont{Raymer}},\ and\ \bibinfo {author}
  {\bibfnamefont{J.~D.}\ \bibnamefont{Bierlein}},\ }%
  \bibfield{journal}{%
  \Doi{10.1364/OL.20.000620}{\bibinfo {journal} {Opt. Lett.}}\ }%
  \textbf{\bibinfo {volume} {20}},\ \bibinfo {pages} {620} (\bibinfo {month}
  {Mar.}\ \bibinfo {year} {1995})%
  \bibAnnoteFile{NoStop}{Anderson95}%
\bibitem{Tanzilli02}%
  \BibitemOpen
  \bibfield{author}{%
  \bibinfo {author} {\bibfnamefont{S.}~\bibnamefont{Tanzilli}}, \bibinfo
  {author} {\bibfnamefont{W.}~\bibnamefont{Tittel}}, \bibinfo {author}
  {\bibfnamefont{H.~D.}\ \bibnamefont{Riedmatten}}, \bibinfo {author}
  {\bibfnamefont{H.}~\bibnamefont{Zbinden}}, \bibinfo {author}
  {\bibfnamefont{P.}~\bibnamefont{Baldi}}, \bibinfo {author}
  {\bibfnamefont{M.}~\bibnamefont{{DeMicheli}}}, \bibinfo {author}
  {\bibfnamefont{D.}~\bibnamefont{Ostrowsky}},\ and\ \bibinfo {author}
  {\bibfnamefont{N.}~\bibnamefont{Gisin}},\ }%
  \bibfield{journal}{%
  \Doi{10.1140/epjd/e20020019}{\bibinfo {journal} {EPJD}}\ }%
  \textbf{\bibinfo {volume} {18}},\ \bibinfo {pages} {155} (\bibinfo {month}
  {Feb.}\ \bibinfo {year} {2002})%
  \bibAnnoteFile{NoStop}{Tanzilli02}%
\bibitem{Kanter02}%
  \BibitemOpen
  \bibfield{author}{%
  \bibinfo {author} {\bibfnamefont{G.}~\bibnamefont{Kanter}}, \bibinfo {author}
  {\bibfnamefont{P.}~\bibnamefont{Kumar}}, \bibinfo {author}
  {\bibfnamefont{R.}~\bibnamefont{Roussev}}, \bibinfo {author}
  {\bibfnamefont{J.}~\bibnamefont{Kurz}}, \bibinfo {author}
  {\bibfnamefont{K.}~\bibnamefont{Parameswaran}},\ and\ \bibinfo {author}
  {\bibfnamefont{M.}~\bibnamefont{Fejer}},\ }%
  \bibfield{journal}{%
  \bibinfo {journal} {Opt. Express}\ }%
  \textbf{\bibinfo {volume} {10}},\ \bibinfo {pages} {177} (\bibinfo {year}
  {2002})%
  \bibAnnoteFile{NoStop}{Kanter02}%
\bibitem{Politi08}%
  \BibitemOpen
  \bibfield{author}{%
  \bibinfo {author} {\bibfnamefont{A.}~\bibnamefont{Politi}}, \bibinfo {author}
  {\bibfnamefont{M.~J.}\ \bibnamefont{Cryan}}, \bibinfo {author}
  {\bibfnamefont{J.~G.}\ \bibnamefont{Rarity}}, \bibinfo {author}
  {\bibfnamefont{S.}~\bibnamefont{Yu}},\ and\ \bibinfo {author}
  {\bibfnamefont{J.~L.}\ \bibnamefont{{O'Brien}}},\ }%
  \bibfield{journal}{%
  \Doi{10.1126/science.1155441}{\bibinfo {journal} {Science}}\ }%
  \textbf{\bibinfo {volume} {320}},\ \bibinfo {pages} {646} (\bibinfo {month}
  {May}\ \bibinfo {year} {2008})%
  \bibAnnoteFile{NoStop}{Politi08}%
\bibitem{Kuzucu08}%
  \BibitemOpen
  \bibfield{author}{%
  \bibinfo {author} {\bibfnamefont{O.}~\bibnamefont{Kuzucu}}, \bibinfo {author}
  {\bibfnamefont{F.~N.~C.}\ \bibnamefont{Wong}}, \bibinfo {author}
  {\bibfnamefont{S.}~\bibnamefont{Kurimura}},\ and\ \bibinfo {author}
  {\bibfnamefont{S.}~\bibnamefont{Tovstonog}},\ }%
  \bibfield{journal}{%
  \Doi{10.1103/PhysRevLett.101.153602}{\bibinfo {journal} {Phys. Rev. Lett.}}\
  }%
  \textbf{\bibinfo {volume} {101}},\ \bibinfo {pages} {153602} (\bibinfo
  {month} {Oct}\ \bibinfo {year} {2008})%
  \bibAnnoteFile{NoStop}{Kuzucu08}%
\bibitem{Law00}%
  \BibitemOpen
  \bibfield{author}{%
  \bibinfo {author} {\bibfnamefont{C.~K.}\ \bibnamefont{Law}}, \bibinfo
  {author} {\bibfnamefont{I.~A.}\ \bibnamefont{Walmsley}},\ and\ \bibinfo
  {author} {\bibfnamefont{J.~H.}\ \bibnamefont{Eberly}},\ }%
  \bibfield{journal}{%
  \Doi{10.1103/PhysRevLett.84.5304}{\bibinfo {journal} {Phys. Rev. Lett.}}\ }%
  \textbf{\bibinfo {volume} {84}},\ \bibinfo {pages} {5304} (\bibinfo {month}
  {Jun}\ \bibinfo {year} {2000})%
  \bibAnnoteFile{NoStop}{Law00}%
\bibitem{Avenhaus09}%
  \BibitemOpen
  \bibfield{author}{%
  \bibinfo {author} {\bibfnamefont{M.}~\bibnamefont{Avenhaus}}, \bibinfo
  {author} {\bibfnamefont{A.}~\bibnamefont{Eckstein}}, \bibinfo {author}
  {\bibfnamefont{P.~J.}\ \bibnamefont{Mosley}},\ and\ \bibinfo {author}
  {\bibfnamefont{C.}~\bibnamefont{Silberhorn}},\ }%
  \bibfield{journal}{%
  \bibinfo {journal} {Opt. Lett.}\ }%
  \textbf{\bibinfo {volume} {34}},\ \bibinfo {pages} {2873} (\bibinfo {year}
  {2009})%
  \bibAnnoteFile{NoStop}{Avenhaus09}%
\bibitem{Avenhaus08}%
  \BibitemOpen
  \bibfield{author}{%
  \bibinfo {author} {\bibfnamefont{M.}~\bibnamefont{Avenhaus}}, \bibinfo
  {author} {\bibfnamefont{H.~B.}\ \bibnamefont{Coldenstrodt-Ronge}}, \bibinfo
  {author} {\bibfnamefont{K.}~\bibnamefont{Laiho}}, \bibinfo {author}
  {\bibfnamefont{W.}~\bibnamefont{Mauerer}}, \bibinfo {author}
  {\bibfnamefont{I.~A.}\ \bibnamefont{Walmsley}},\ and\ \bibinfo {author}
  {\bibfnamefont{C.}~\bibnamefont{Silberhorn}},\ }%
  \bibfield{journal}{%
  \Doi{10.1103/PhysRevLett.101.053601}{\bibinfo {journal} {Phys. Rev. Lett.}}\
  }%
  \textbf{\bibinfo {volume} {101}},\ \bibinfo {pages} {053601} (\bibinfo
  {month} {Aug}\ \bibinfo {year} {2008})%
  \bibAnnoteFile{NoStop}{Avenhaus08}%
\bibitem{Fiorentino07}%
  \BibitemOpen
  \bibfield{author}{%
  \bibinfo {author} {\bibfnamefont{M.}~\bibnamefont{Fiorentino}}, \bibinfo
  {author} {\bibfnamefont{S.~M.}\ \bibnamefont{Spillane}}, \bibinfo {author}
  {\bibfnamefont{R.~G.}\ \bibnamefont{Beausoleil}}, \bibinfo {author}
  {\bibfnamefont{T.~D.}\ \bibnamefont{Roberts}}, \bibinfo {author}
  {\bibfnamefont{P.}~\bibnamefont{Battle}},\ and\ \bibinfo {author}
  {\bibfnamefont{M.~W.}\ \bibnamefont{Munro}},\ }%
  \bibfield{journal}{%
  \bibinfo {journal} {Opt. Express}\ }%
  \textbf{\bibinfo {volume} {15}},\ \bibinfo {pages} {7479} (\bibinfo {year}
  {2007})%
  \bibAnnoteFile{NoStop}{Fiorentino07}%
\bibitem{Gottesmann01}%
  \BibitemOpen
  \bibfield{author}{%
  \bibinfo {author} {\bibfnamefont{D.}~\bibnamefont{Gottesman}}\ and\ \bibinfo
  {author} {\bibfnamefont{J.}~\bibnamefont{Preskill}},\ }%
  \bibfield{journal}{%
  \Doi{10.1103/PhysRevA.63.022309}{\bibinfo {journal} {Phys. Rev. A}}\ }%
  \textbf{\bibinfo {volume} {63}},\ \bibinfo {pages} {022309} (\bibinfo {month}
  {Jan}\ \bibinfo {year} {2001})%
  \bibAnnoteFile{NoStop}{Gottesmann01}%
\bibitem{Furusawa98}%
  \BibitemOpen
  \bibfield{author}{%
  \bibinfo {author} {\bibfnamefont{A.}~\bibnamefont{Furusawa}}, \bibinfo
  {author} {\bibfnamefont{J.~L.}\ \bibnamefont{S{\o}rensen}}, \bibinfo {author}
  {\bibfnamefont{S.~L.}\ \bibnamefont{Braunstein}}, \bibinfo {author}
  {\bibfnamefont{C.~A.}\ \bibnamefont{Fuchs}}, \bibinfo {author}
  {\bibfnamefont{H.~J.}\ \bibnamefont{Kimble}},\ and\ \bibinfo {author}
  {\bibfnamefont{E.~S.}\ \bibnamefont{Polzik}},\ }%
  \bibfield{journal}{%
  \Doi{10.1126/science.282.5389.706}{\bibinfo {journal} {Science}}\ }%
  \textbf{\bibinfo {volume} {282}},\ \bibinfo {pages} {706} (\bibinfo {month}
  {Oct.}\ \bibinfo {year} {1998})%
  \bibAnnoteFile{NoStop}{Furusawa98}%
\bibitem{Bowen04}%
  \BibitemOpen
  \bibfield{author}{%
  \bibinfo {author} {\bibfnamefont{W.~P.}\ \bibnamefont{Bowen}}, \bibinfo
  {author} {\bibfnamefont{N.}~\bibnamefont{Treps}}, \bibinfo {author}
  {\bibfnamefont{B.~C.}\ \bibnamefont{Buchler}}, \bibinfo {author}
  {\bibfnamefont{R.}~\bibnamefont{Schnabel}}, \bibinfo {author}
  {\bibfnamefont{T.~C.}\ \bibnamefont{Ralph}}, \bibinfo {author}
  {\bibfnamefont{H.-A.}\ \bibnamefont{Bachor}}, \bibinfo {author}
  {\bibfnamefont{T.}~\bibnamefont{Symul}},\ and\ \bibinfo {author}
  {\bibfnamefont{P.~K.}\ \bibnamefont{Lam}},\ }%
  \bibfield{journal}{%
  \Doi{10.1103/PhysRevA.67.032302}{\bibinfo {journal} {Phys. Rev. A}}\ }%
  \textbf{\bibinfo {volume} {67}},\ \bibinfo {pages} {032302} (\bibinfo {month}
  {Mar}\ \bibinfo {year} {2003})%
  \bibAnnoteFile{NoStop}{Bowen04}%
\end{thebibliography}


%

\end{document}